\documentclass[prl,twocolumn]{revtex4}
\usepackage{epsfig,amssymb,amsmath,graphicx,enumerate,color}
\pdfoutput = 1

\begin{document}

\title{Structured optical receivers to attain superadditive capacity and the Holevo limit}

\author{Saikat Guha}
\affiliation{Disruptive Information Processing Technologies group, Raytheon BBN Technologies, Cambridge, MA 02138, USA}

\begin{abstract}
When classical information is sent over a quantum channel, attaining the ultimate limit to channel capacity requires the receiver to make joint measurements over long codeword blocks. For a pure-state channel, we construct a receiver that can attain the ultimate capacity by applying a single-shot unitary transformation on the received quantum codeword followed by simultaneous (but separable) projective measurements on the single-modulation-symbol state spaces. We study the ultimate limits of photon-information-efficient communications on a lossy bosonic channel. Based on our general results for the pure-state quantum channel, we show some of the first concrete examples of codes and structured joint-detection optical receivers that can achieve fundamentally higher (superadditive) channel capacity than conventional receivers that detect each modulation symbol individually.
\end{abstract}

\maketitle

When the modulation alphabet of a communication channel are quantum states, the Holevo limit is an upper bound to the Shannon capacity of the physical channel paired with any receiver measurement. Even though the Holevo limit is an achievable capacity, the receiver in general must make joint ({\em collective}) measurements over long codeword blocks---measurements that can't be realized by detecting single modulation symbols followed by classical post processing. This phenomenon of a joint-detection receiver (JDR) being able to yield higher capacity than any single-symbol receiver measurement, is often termed as {\em superadditivity} of capacity.

For the lossy bosonic channel, a coherent-state modulation suffices to attain the Holevo capacity, i.e., non-classical transmitted states do not yield any additional capacity~\cite{Gio04}. Hausladen et. al.'s square-root-measurement (SRM)~\cite{Hau96}, which in general is a positive operator-valued measure (POVM), applied to a random code gives us the mathematical construct of a receiver that can achieve the Holevo limit. Lloyd et. al.~\cite{Llo10} recently showed a receiver that can attain the Holevo capacity of any quantum channel by making a sequence of ``yes/no" projective measurements on a random codebook. Sasaki et. al.~\cite{Sas98}, in a series of papers, showed several examples of superadditive capacity using pure-state alphabets and the SRM. However, the key practical questions that remain unanswered are how to design modulation formats, channel codes, and most importantly, structured realizations of Holevo-capacity-approaching JDRs. 

In this paper, (i) we show that the Holevo limit of a pure-state channel is attained by a projective measurement, which can be implemented by a unitary operation on the quantum codeword followed by separable projective measurements on the single-modulation-symbol subspaces, (ii) we translate our result into an optimal receiver for the lossy bosonic channel, and (iii) we show concrete examples of codes and receivers that yield superadditive capacity for optical binary-phase-shift keying (BPSK) signaling at low photon numbers. These, we believe, are the first receiver realizations that can exhibit superaddivity, and can be tested using laboratory optics.

\begin{figure}
\centering
\includegraphics[width=0.9\columnwidth]{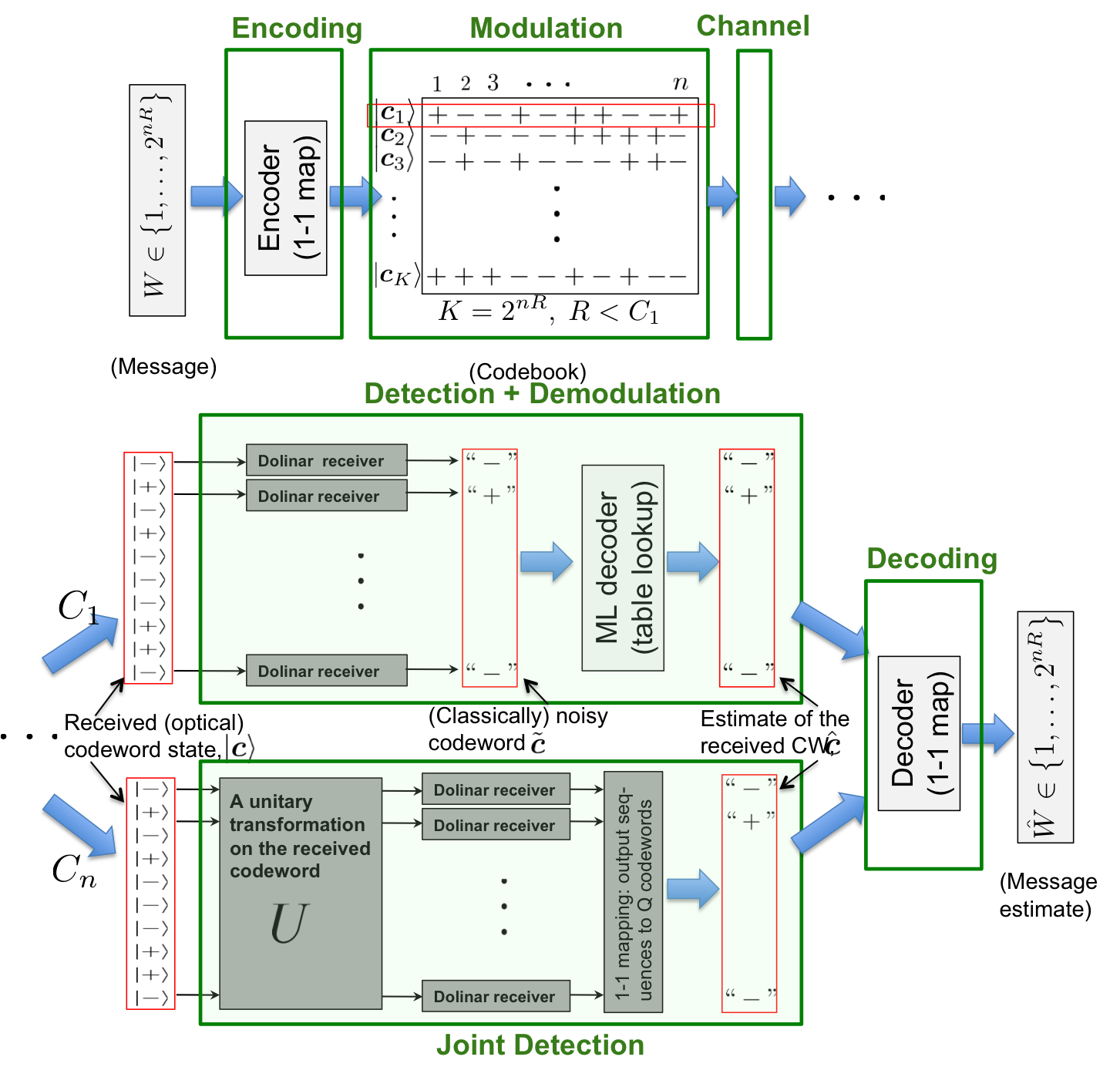}
\vspace{-10pt}
\caption{Classical communication system, shown here for a BPSK alphabet. If the receiver uses symbol-by-symbol detection, maximum capacity $= C_1$ bits/symbol. If the Detection$+$Demodulation block is replaced by a general $n$-input $n$-output quantum measurement, maximum capacity $= C_n$ bits/symbol. Superadditivity: $C_\infty > C_n > C_1$, where $C_\infty$ is the Holevo limit. The joint-detection structure shown achieves Holevo limit for BPSK over a lossy bosonic channel.}
\label{fig:classicalsystem}
\end{figure}

{\bf Attaining Holevo limit of a pure-state channel.} We encode classical information using a $Q$-ary modulation alphabet of non-orthogonal pure-state {\em symbols} in $S \equiv \left\{|\psi_1\rangle, \ldots, |\psi_Q\rangle\right\}$. Each {\em channel use} constitutes sending one symbol. We assume that the channel preserves the purity of $S$, thus taking the states $\left\{|\psi_q\rangle\right\}$ to be those at the receiver. The only source of noise is the physical detection of the states. Assume that the receiver detects each symbol one at a time. Channel capacity is given by the maximum of the single-symbol mutual information, 
\begin{equation}
C_1 = \max_{\left\{p_i\right\}}\max_{\left\{{\hat \Pi}_j^{(1)}\right\}}I_1\left(\left\{p_i\right\}, \left\{{\hat \Pi}_j^{(1)}\right\}\right) {\text{ bits/symbol}},
\label{eq:C1definition}
\end{equation}
where the maximum is taken over priors $\left\{p_i\right\}$ over the alphabet and a set of POVM operators $\left\{{\hat \Pi}_j^{(1)}\right\}$, $1 \le j \le J$ on the single-symbol state-space. The measurement of each symbol produces one of $J$ possible outcomes, with conditional probability $P(j|i) = \langle{\psi_i}|{\hat \Pi}_j^{(1)}|\psi_i\rangle$. To achieve reliable communications at a rate close to $C_1$, forward error-correction will need to be applied on the discrete memoryless channel with transition probabilities $P(j|i)$. In other words, for any rate $R < C_1$, there exists a sequence of codebooks ${\cal C}_n$ with $K = 2^{nR}$ codewords $|{\boldsymbol c}_k\rangle$, $1 \le k \le K$, each codeword being an $n$-symbol tensor product of states in $S$, and a decoding rule, such that the average probability of decoding error (guessing the wrong codeword), ${\bar P}_e^{(n)} = 1-\frac1K\sum_{k=1}^K{\rm Pr}({\hat k}=k) \to 0$, as $n \to \infty$. In this `Shannon' setting, optimal decoding is a maximum likelihood (ML) decision, which can in principle be pre-computed as a table lookup (see Fig.~\ref{fig:classicalsystem}).
We define $C_n$ as the maximum capacity achievable with measurements that jointly detect up to $n$ symbols. The fact that joint detection allows for $C_{n+m} > C_n + C_m$, (or $C_n > C_1$) is referred to as {\em superadditivity} of capacity. The Holevo-Schumacher-Westmorland (HSW) theorem says,
\begin{equation}
C_\infty \equiv \max_{\left\{p_i\right\}}S\left(\sum_ip_i|\psi_i\rangle\langle{\psi_i}|\right) = \lim_{n\to \infty}C_n,
\label{eq:HSWpure}
\end{equation}
the Holevo bound, is the ultimate capacity limit, where $S({\hat \rho}) = -{\rm Tr}{\hat \rho}\log_2{\hat \rho}$ is the von Neumann entropy, and that $C_\infty$ is achievable with joint detection over long codeword blocks. Calculating $C_\infty$ however, doesn't require the knowledge of the optimal receiver measurement. In other words, if we replaced the detection and demodulation stages in Fig.~\ref{fig:classicalsystem} by one giant quantum measurement, then for any rate $R < C_\infty$, there exists a sequence of codebooks ${\cal C}_n$ with $K = 2^{nR}$ codewords $|{\boldsymbol c}_k\rangle$, $1 \le k \le K$, and an $n$-input $n$-output POVM over the $n$-symbol state-space $\left\{{\hat \Pi}_k^{(n)}\right\}$, $1 \le k \le K$, such that the average probability of decoding error, ${\bar P}_e^{(n)} = 1-\frac1K\sum_{k=1}^K\langle{\boldsymbol c}_k|{\hat \Pi}_k^{(n)}|{\boldsymbol c}_k\rangle \to 0$, as $n \to \infty$.

{\it Theorem 1:} For a pure-state channel, a projective measurement can attain $C_\infty$, and can be implemented as a unitary transformation on the codeword followed by a parallel set of separable single-symbol measurements.

{\it Proof:} Consider a codebook ${\cal C}$ with $K = 2^{nR}$ codewords $|{\boldsymbol c}_k\rangle$, $1 \le k \le K$, each codeword being an $n$-symbol tensor product of states in $S$. The SRM (which in general is a POVM) on a random codebook can achieve the Holevo capacity~\cite{Hau96}. However, Helstrom showed that the minimum probability of error (MPE) measurement for discriminating $K$ pure states with the least average probability of error is a $K$-element projective measurement on the span of those $K$ states~\cite{Hel76}. By definition, the MPE measurement on $\cal C$ must achieve a lower average probability of error in discriminating the codewords in ${\cal C}$ than the SRM. Hence, the MPE measurement is capacity achieving. In other words, given any {\em reliable communication} threshold on the decoding error rate, $P_{\rm th}$, there exists a codebook $\cal C$ of a long enough length $n$, such that the MPE measurement on $\cal C$ described by the projectors $\left\{{\boldsymbol {\hat \Pi}}_{\rm MPE}^{(n)}\right\}$ with ${\hat \Pi}_{{\rm MPE},k}^{(n)}=|{\boldsymbol w}_k\rangle\langle{\boldsymbol w}_k|$ (where $\left\{|{\boldsymbol w}_1\rangle, \ldots, |{\boldsymbol w}_K\rangle\right\}$ form a complete ortho-normal (CON) basis for ${\rm span}({\cal C})$), can attain a probability of decoding error $P_e^{(n)} = 1-(1/K)\sum_{k=1}^K|\langle{\boldsymbol c}_k|{\boldsymbol w}_k\rangle|^2 \le P_{\rm th}$. Now, let us define the MPE measurement on the states in $S$ as ${\hat \Pi}_{\rm MPE}^{(1)} \equiv \left\{|m_q\rangle\langle{m_q}|\right\}$, where $\left\{|m_1\rangle, \ldots, |m_Q\rangle\right\}$ is a CON basis of the single-symbol state space ${\cal H}_1 \equiv {\rm span}(S)$. Let us define the Kronecker-product CON basis of the $n$-symbol state space ${\cal H}_n = {\cal H}_1^{\otimes n}$, $M \equiv \left\{|{\boldsymbol m}_1\rangle, \ldots, |{\boldsymbol m}_{Q^n}\rangle\right\}$, where $|{\boldsymbol m}_{\boldsymbol q}\rangle = |m_{q_1}\rangle|m_{q_2}\rangle\ldots|m_{q_n}\rangle$, where $q_k \in [1, \ldots, Q]$, $\forall k$. Let us define another CON basis of the $n$-symbol state space ${\cal H}_n$ by extending $\left\{|{\boldsymbol w}_1\rangle, \ldots, |{\boldsymbol w}_K\rangle\right\}$---the vectors describing the MPE measurement on $\cal C$---as $W \equiv \left\{|{\boldsymbol w}_1\rangle, \ldots, |{\boldsymbol w}_{K}\rangle, |{\boldsymbol w}_{K+1}\rangle, \ldots, |{\boldsymbol w}_{Q^n}\rangle\right\}$. Finally, let us extend the codebook $\cal C$ into the set of {\em all} $Q^n$ length-$n$ sequences of modulation symbols
$
{\cal C}^E \equiv \left\{|{\boldsymbol c}_1\rangle, \ldots, |{\boldsymbol c}_K\rangle, |{\boldsymbol c}_{K+1}\rangle, \ldots, |{\boldsymbol c}_{Q^n}\rangle\right\},
$
and express each $|{\boldsymbol c}_k\rangle$ in both CON bases $M$ and $W$,
$
|{\boldsymbol c}_k\rangle = \sum_{j=1}^{Q^n}\langle{\boldsymbol m}_j|{\boldsymbol c}_k\rangle|{\boldsymbol m}_j\rangle 
$,
$
|{\boldsymbol c}_k\rangle = \sum_{j=1}^{Q^n}\langle{\boldsymbol w}_j|{\boldsymbol c}_k\rangle|{\boldsymbol w}_j\rangle
$,
where $\langle{\boldsymbol m}_j|{\boldsymbol c}_k\rangle \equiv (U_M)_{kj}$ and $\langle{\boldsymbol w}_j|{\boldsymbol c}_k\rangle \equiv (U_W)_{kj}$ are $(k,j)^{\rm th}$ elements of the unitary matrices $U_M$ and $U_W$. The bases $M$ and $W$ are unitarily equivalent. Thus, the MPE measurement on $\cal C$ is equivalent to first applying a unitary $U$ on the codeword followed by the projective measurement described by $M$, which is essentially $n$ parallel (and separable) MPE measurements $\left\{{\hat \Pi}_{\rm MPE}^{(1)}\right\}$, on the single-symbol subspaces. The unitary $U$ is given by (in the basis $M$),
\begin{equation}
U = \sum_{k=1}^{Q^n}\sum_{j=1}^{Q^n}u_{jk}^*|{\boldsymbol m}_j\rangle\langle{\boldsymbol m}_k|; \quad u_{jk} = \left(U_W^{-1}U_M\right)_{jk}.
\end{equation}

\begin{figure}
\centering
\includegraphics[width=0.7\columnwidth]{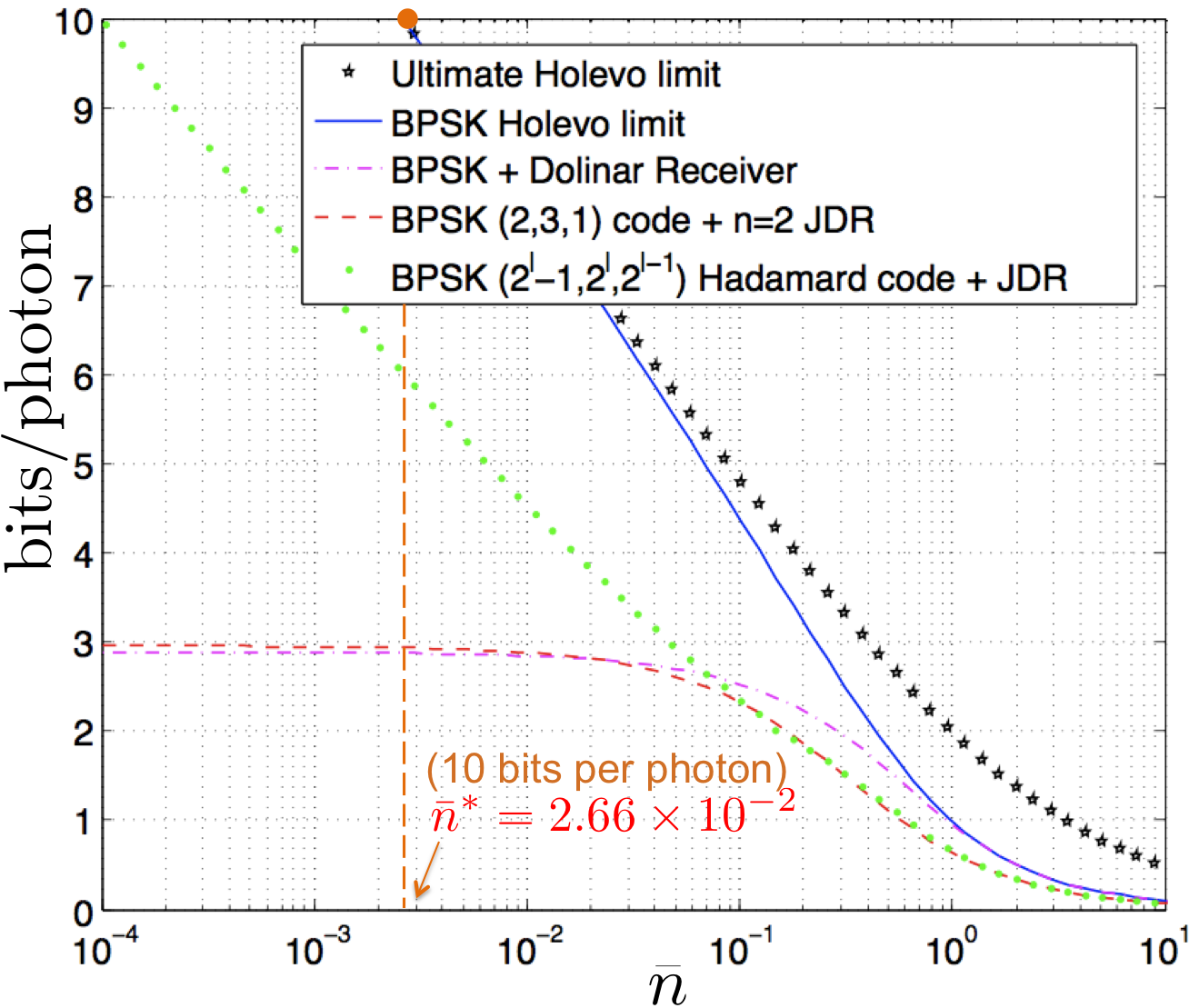}
\vspace{-10pt}
\caption{\small{Photon information efficiency (bits per received photon) as a function of mean photon number per mode, $\bar n$.}}
\label{fig:PIE_vs_nbar}
\end{figure}
{\bf Superadditive optical receivers.} Consider a single-mode lossy bosonic channel (such as a far-field single-spatial-mode free-space-optical (FSO) channel), where data is modulated using a succession of pulses (orthogonal temporal modes) with mean received photon number $\bar n$ per mode, where each temporal mode (pulse) carries one modulation symbol. The Holevo capacity is given by, 
$
C_{\rm ult}({\bar n}) = g({\bar n})=(1+{\bar n})\log_2(1+{\bar n})-{\bar n}\log_2{\bar n} {\text{ bits/symbol}},
$
which is attained using a coherent-state modulation, i.e., non-classical modulation states can't get any higher capacity~\cite{Gio04}. Since pure loss preserves coherent states (with linear amplitude attenuation), it suffices to define capacity as a function of the mean photon number per {\em received} mode $\bar n$, and the results derived above for a pure-state channel applies. Achieving the Holevo limit requires an optimal codebook and joint measurement on long codeword blocks. At high $\bar n$, symbol-by-symbol heterodyne detection asymptotically achieves $C_{\rm ult}({\bar n})$. The low photon number (${\bar n} \ll 1$) regime is more interesting for far-field FSO communications, where the joint-detection gain is most pronounced. 

In Fig.~\ref{fig:PIE_vs_nbar}, we show the photon information efficiency (PIE), the number of bits that can be reliably decoded per received photon, as a function of $\bar n$~\footnote{PIE is a more useful metric to communication engineers than channel capacity $C(\bar n)$. It translates readily into a tradeoff between {\em photon efficiency}, $C({\bar n})/{\bar n}$ bits/photon, and {\em spectral efficiency}, $C({\bar n})$ bits/sec/Hz.}. There is no fundamental upper bound to the PIE; however, higher PIE necessitates lower $\bar n$. Furthermore, binary modulation and coding is sufficient to meet the Holevo limit at low $\bar n$. Specifically, the binary-phase-shift keying (BPSK) alphabet $S_1 \equiv \left\{|\alpha\rangle, |-\alpha\rangle\right\}$, $|\alpha|^2 = {\bar n}$, is the Holevo-optimal binary modulation at ${\bar n} \ll 1$. Dolinar proposed a structured receiver that realizes the binary MPE projective measurement on an a pair of coherent states using single photon detection and coherent optical feedback~\cite{Dol76}. If the Dolinar receiver (DR) is used to detect each symbol, the BPSK channel is reduced to a classical binary symmetric channel (BSC) with capacity $C_1 = 1-H(q)$ bits/symbol, $q = [1-\sqrt{1-e^{-4{\bar n}}}]/2$. This is the maximum achievable capacity when the receiver detects each symbol individually, which includes all conventional (direct-detection and coherent-detection) receivers. The PIE $C_1({\bar n})/{\bar n}$ caps out at $2/\ln 2 \approx 2.89$ bits/photon at ${\bar n} \ll 1$. Closed-form expressions and scaling behavior of $C_n$, the maximum capacity achievable with measurements that jointly detect up to $n$ symbols, for $n \ge 2$ are not known. However, the Holevo limit of BPSK, $C_\infty(\bar n) = H([1+e^{-2{\bar n}}]/2)$, can be calculated using Eq.~\eqref{eq:HSWpure}. Good codes and JDRs will be needed to bridge the huge gap between the PIEs $C_1(\bar n)/{\bar n}$ and $C_\infty({\bar n})/{\bar n}$, shown in Fig.~\ref{fig:PIE_vs_nbar}. It is interesting to reflect on the point shown by the orange circle (at $10$ bits/photon) in Fig.~\ref{fig:PIE_vs_nbar}, which says that for a $1.55$$\mu$m far-field FSO system operating at $1$ GHz modulation bandwidth, the laws of physics permit reliable communications at $0.266$ Gbps with only $3.4$ pW of average (and peak) received optical power! 

\begin{figure}
\centering
\includegraphics[width=0.85\columnwidth]{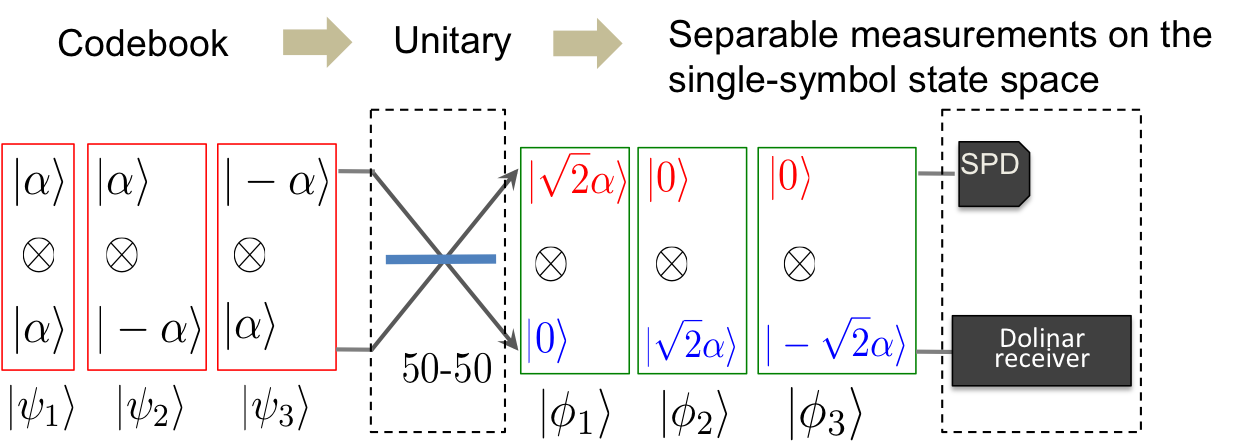}
\vspace{-10pt}
\caption{\small{A two-symbol JDR that attains $\approx 2.5\%$ higher capacity for BPSK than the best single-symbol (Dolinar) receiver.}}
\label{fig:measurements}
\end{figure}
{\em A two-symbol superadditive JDR---} Some examples of superadditive codes and joint measurements have been reported~\cite{Fuc97, Sas98}, but no structured receiver designs. An ensemble (a $(2,3,1)$ inner code~\footnote{An $(n,K,d)$ code has $K$ length-$n$ codewords, such that the minimum Hamming distance between any pair of codewords is $d$. The {\em code rate} is $R = \log_2K/n$.}) containing three of the four $2$-symbol BPSK states, $S_2 \equiv \left\{|\alpha\rangle|\alpha\rangle, |\alpha\rangle|-\alpha\rangle, |-\alpha\rangle|\alpha\rangle \right\}$, with priors $(1-2p, p, p)$, $0 \le p \le 0.5$, can attain, with the best $3$-element projective measurement in ${\rm span(S_2)}$, up to $\approx 2.8\%$ higher capacity that $C_1$~\cite{Fuc97}. Since this is a Shannon capacity result, a classical outer code with codewords comprising of sequences of states from $S_2$ will be needed to achieve this capacity $I_2 > C_1$. Using the MPE measurement on $S_2$ (which can be analytically calculated~\cite{Hel76}, unlike the numerically optimized projections in~\cite{Fuc97}), $I_2/C_1 \approx 1.0266$ can be obtained. A receiver that involves a unitary operation on the $[2,3,1]$ code (a beamsplitter) followed by two separable single-symbol measurements (in this case, a single-photon detector (SPD), and a DR) (see Fig.~\ref{fig:measurements}), can attain $I_2/C_1 \approx 1.0249$ (see Fig.~\ref{fig:PIE_vs_nbar}). Its likely that none of these projective measurements on $S_2$ attain $C_2$, since the single-shot measurement that maximizes the accessible information in $S_2$ could in general be a $6$-element POVM~\cite{Sho00}. However, {\em Theorem 1} proves that as the size of the inner code $n \to \infty$, a projective measurement on the codebook that involves an $n$-mode unitary followed by a DR-array is capable of attaining $C_\infty$, without any additional outer code (see Fig.~\ref{fig:classicalsystem}). 

\begin{figure}[h]
\centering
\includegraphics[width=0.9\columnwidth]{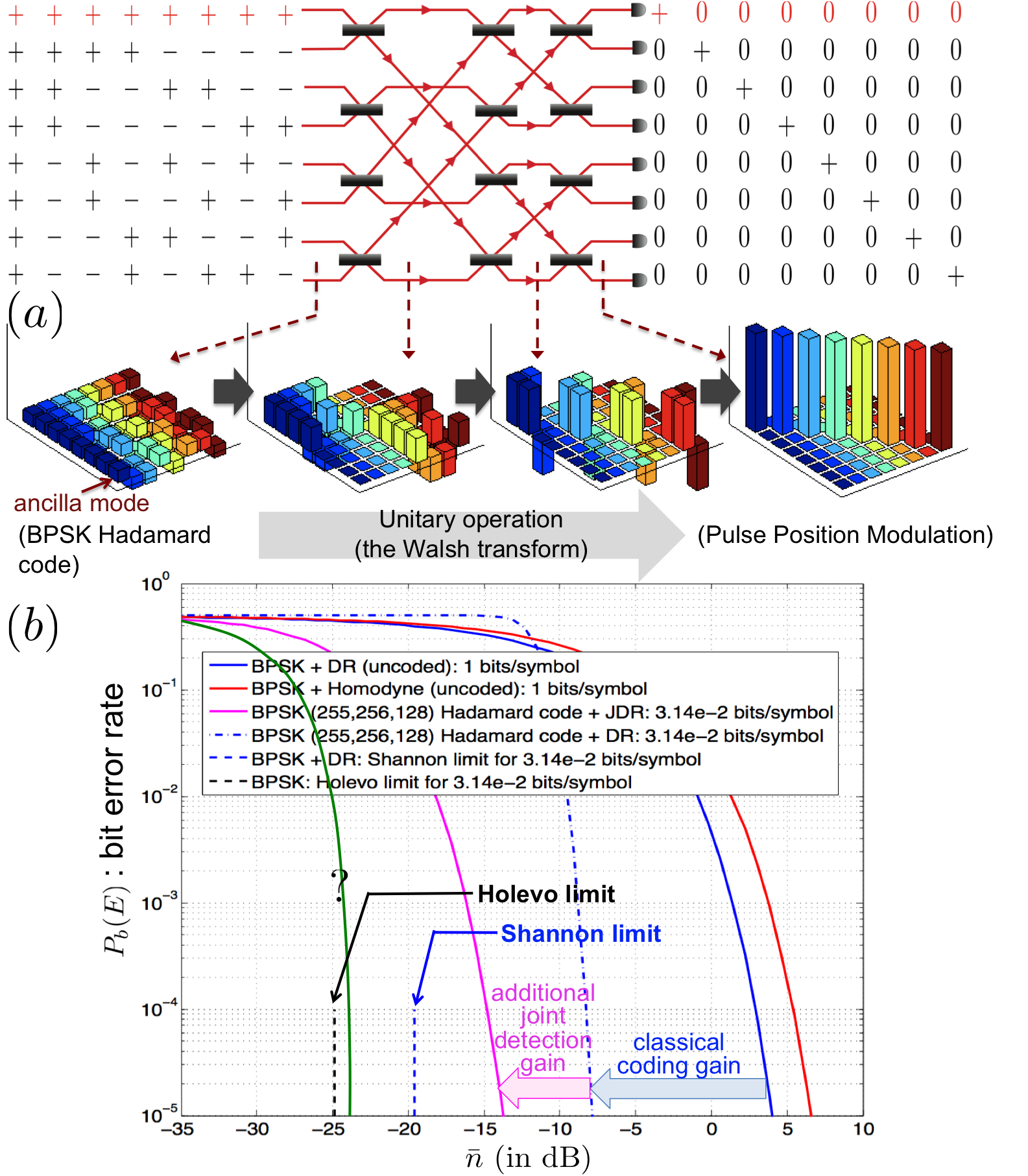}
\vspace{-10pt}
\caption{\small{(a) The BPSK $(7,8,4)$ Hadamard code is unitarily equivalent to the $(8,8,4)$ pulse-position-modulation (PPM) code via a Walsh transform built using twelve 50-50 beamsplitters. (b) Bit error rate plotted as a function of $\bar n$.}}
\label{fig:HadamardJDR}
\end{figure}

{\em An $n$-symbol superadditive JDR---} A $(2^l-1,2^l,2^{l-1})$ BPSK Hadamard code, with $\bar n$-mean-photons BPSK symbols, has the same geometry (mutual inner-products) and thus is unitarily equivalent to the $(2^l,2^l,2^{l-1})$ pulse-position-modulation (PPM) code with $2^l{\bar n}$-mean-photon PPM pulses. The former is slightly {\em space-efficient}, since it achieves the same equidistant distance profile, but with one less symbol. Consider a BPSK Hadamard code detected by a $2^l$-mode unitary transformation (with one ancilla mode prepared locally at the receiver, in the $|\alpha\rangle$ state) built using $(n\log_2n)/2$ 50-50 beamsplitters arranged in a fast-fourier-transform (FFT) butterfly circuit, followed by a separable $n = 2^l$-element SPD-array, as shown (for $n = 8$) in Fig.~\ref{fig:HadamardJDR}(a). The beamsplitter array is reminiscent of the fast Walsh Hadamard transform, if we recall that each 50-50 beamsplitter implements an order-$2$ FFT. The beamsplitters `unravel' the BPSK codebook into a PPM codebook, separating out the photon energies in spatially separated bins, making possible discriminating the codewords using a SPD array. This receiver design is a more `natural' choice for spatial modulation, across, say orthogonal spatial modes of a near-field FSO channel. The ancilla mode at the receiver necessitates a local oscillator phase locked to the received pulses, which is hard to implement. Since the number of ancilla modes doesn't scale with the size of the code, we can append the ancilla mode to the transmitted codeword, so that the received ancilla can serve as a pilot tone for our interferometric receiver. The Shannon capacity of this code-JDR pair---allowing coding over the {\em erasure outcome} (no clicks registered at any SPD element)---is $I_n(\bar n) = (\log_2K/K)(1-\exp(-2d{\bar n}))$ bits/symbol. In Fig.~\ref{fig:PIE_vs_nbar}, we plot the envelope, $\max_nI_n(\bar n)/{\bar n}$ (the green dotted plot), as a function of $\bar n$. This JDR not only attains a {\em much} higher superadditive gain than the $n=2$ JDR we describe above, it doesn't need phase tracking and coherent optical feedback like the DR. Note that $n$-ary PPM signaling also achieves the capacity $I_n(\bar n)$ with a SPD receiver, albeit with a much higher ($\times 2^l$) peak power as compared to BPSK. However, {\em Theorem 1} says that the receiver construct shown in Fig.~\ref{fig:classicalsystem} is capable of bridging the rest of the gap to the Holevo limit (i.e., the blue plot in Fig.~\ref{fig:PIE_vs_nbar}) in conjunction with an optimal BPSK code. {\em Theorem 1} applies readily to any higher-order modulation format ($Q>2$, required to achieve capacity at higher $\bar n$), where the DR in the JDR structure must be replaced by an extension of the DR that discriminates $Q$ modulation symbols at their MPE limit (not known yet). Finally, note that in the proof of {\em Theorem 1}, the MPE measurement on the single-symbol space ${\hat \Pi}_{\rm MPE}^{(1)}$, was just a convenient choice; any other projective measurement would've worked. In the optimal JDR for BPSK therefore, the DRs can be replaced by an array of Kennedy receivers (which applies a coherent shift by $-\alpha$ followed by SPD---a lot simpler than the DR), since it also performs a projective measurement on ${\rm span}(S_1)$. In Fig.~\ref{fig:HadamardJDR}(b), we plot the bit error rates $P_b(E)$ as a function of $\bar n$ for uncoded BPSK, and the $[255,256,128]$ BPSK Hadamard code, when detected using a symbol-by-symbol DR and our structured JDR, respectively. The {\em coding gain} now has two components, a (classical) coding gain, and an additional {\em joint-detection gain}. 

A great deal is known about binary codes that achieve low bit error rates on the BSC at $\bar n$ very close to the Shannon limit~\cite{For66}. It would be interesting to see how close to the Holevo limit can these same codes perform, when paired with their respective quantum MPE measurements. It will be useful to design codes with symmetries that allow them to approach Holevo capacity, with the unitary $U$ of the JDR in Fig.~\ref{fig:classicalsystem} realizable via a simple network of beamsplitters, phase shifters, two-mode squeezers, and Kerr non-linearities (which form a universal set for realizing an arbitrary multimode bosonic unitary~\cite{Sef10}) along with a low-complexity outer code, if at all. The fields of information and coding theory have had a unique history. Even though many of its ultimate limits were determined in Shannon's founding paper~\cite{Sha48}, it took generations of magnificent coding theory research, to ultimately find practical capacity-approaching codes. Even though realizing reliable communications on an optical channel close to the Holevo limit might take a while, it certainly does seem to be in the visible horizon.

\begin{acknowledgments}
This work was supported by the DARPA Information in a Photon program, contract\#HR0011-10-C-0159. Discussions with Profs. Jeffrey H. Shapiro, Seth Lloyd, and Lizhong Zheng, MIT, Dr. Zachary Dutton, BBN and Dr. Mark Neifeld, DARPA are gratefully acknowledged.
\end{acknowledgments}

\end{document}